\documentclass[11pt]{article}
\usepackage{moriond}

\def\Journal#1#2#3#4{{#1} {\bf #2}, #3 (#4)}

\def\PLB{{\em Phys. Lett.}  B}
\def\PRL{\em Phys. Rev. Lett.}
\def\PRD{{\em Phys. Rev.} D}

\def\be{\begin{equation}}
\def\ee{\end{equation}}
\def\bea{\begin{eqnarray}}
\def\eea{\end{eqnarray}}

\newcommand{\Photo}{\includegraphics[height=35mm]{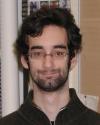}}

\begin{document}
\vspace*{4cm}
\title{SEARCH FOR NEW HEAVY NEUTRAL BOSONS DECAYING INTO A DILEPTON PAIR WITH THE CMS DETECTOR AT $\sqrt{s} =$ 8 TEV}
\author{L. THOMAS,\\FOR THE CMS COLLABORATION }

\address{Interuniversity Institute for High Energies and Universit\'e Libre de Bruxelles, ULB, \\ 
Boulevard du Triomphe, 2, 1050 Brussels, Belgium \\~}

\maketitle\abstracts{Several theories beyond the Standard Model predict the existence of new heavy neutral bosons. 
Such particles could be produced in significant amounts at the LHC and their decay into a dilepton pair provides a clean signature with low background contamination. 
The results of the analysis of the whole 2012 dataset collected by the CMS experiment at a center of mass energy of 8 TeV are presented. 
No evidence for new physics is seen and upper limits on the cross section production of such particles are extracted. 
These results can be turned into lower limits on the mass of the heavy bosons, reaching values well above 2 TeV/c$^2$ for many models.}

\section{Introduction}
Many scenarios beyond the Standard Model (e.g. Grand Unification Theories~\cite{Leike}, models with extra spatial dimensions~\cite{RS}) involve new neutral bosons with masses in the TeV range.
Thanks to its high energy proton beams and high integrated luminosity, the LHC could produce such particles in significant amounts to allow to detect them. 
The dilepton decay channel presents the advantage of a clear signature with small background contamination coming essentially from the Drell-Yan process~\cite{DY,CMSDY} that is the production of a dilepton pair through a photon or a Z boson ($q\bar{q}\rightarrow l\bar{l}$). 
The existence of these new particles would appear as a peak in the dilepton invariant mass spectrum~\cite{PAS,Paper}. 
The dielectron and dimuon channels present complementary features : the former benefits from an excellent mass resolution ($\sim$1-2\%) while the latter is less contaminated by fakes. 
In the following, we focus on the dielectron channel. 
The whole dataset recorded by CMS at a center of mass energy of 8 TeV in year 2012 is analysed. 
It corresponds to an integrated luminosity of about 20 fb$^{-1}$.

\section{The CMS detector}
The central feature of the Compact Muon Solenoid (CMS)\cite{CMS} apparatus is a superconducting solenoid of 6 m internal diameter, providing a magnetic field of 3.8 T. 
Within the superconducting solenoid volume are a silicon pixel and strip tracker, a lead tungstate crystal electromagnetic calorimeter (ECAL), and a brass/scintillator hadron calorimeter (HCAL). 
Muons are measured in gas-ionization detectors embedded in the steel return yoke outside the solenoid. 
Extensive forward calorimetry complements the coverage provided by the barrel and endcap detectors. 

\section{Event Selection}
\begin{figure}[h]
\centerline{\includegraphics[height=5cm]{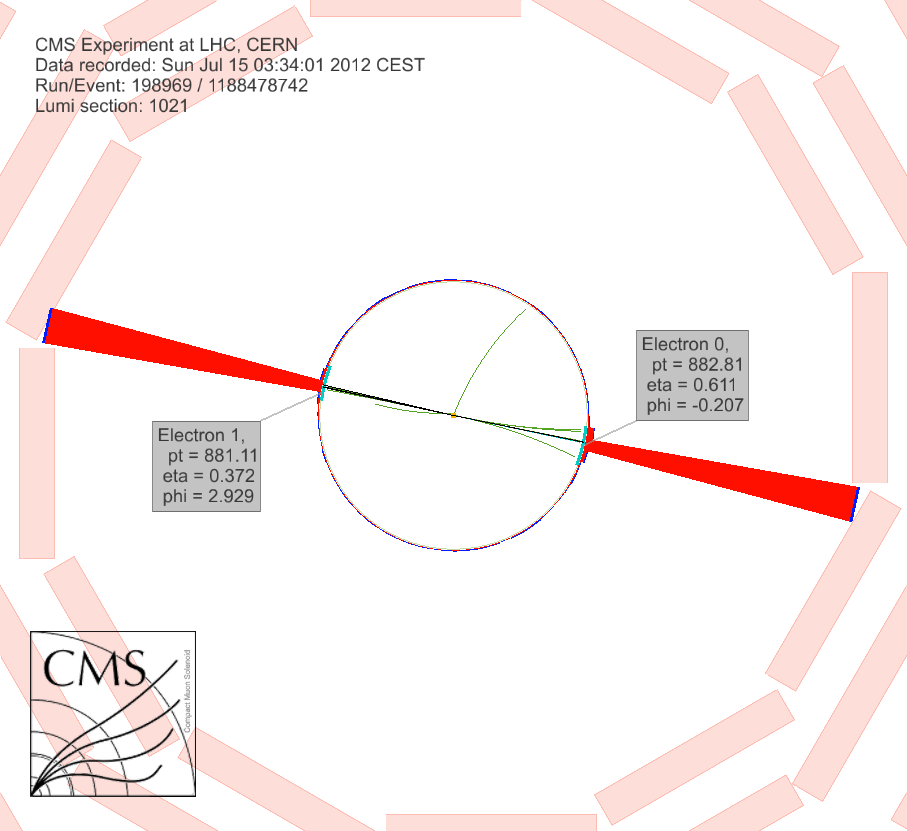}\includegraphics[height=5.1cm]{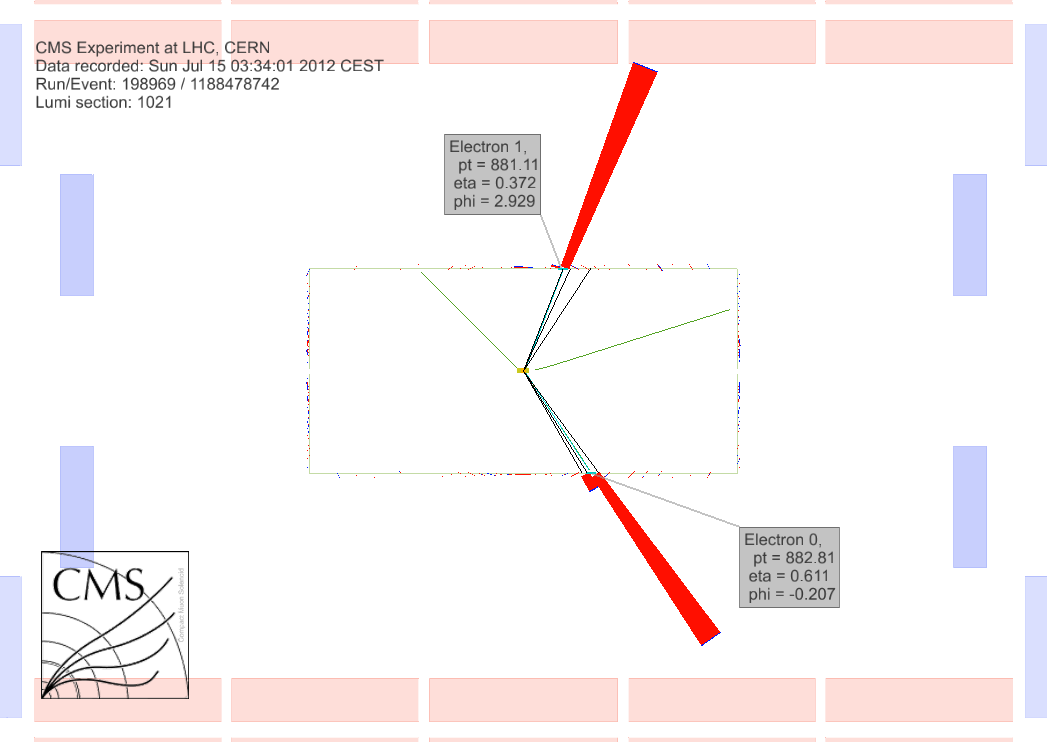}}
\caption{Transverse (left) and longitudinal (right) views of the highest dielectron mass event recorded by the CMS detector in 2012. The dielectron mass is 1776 GeV/c$^2$. }
\label{fig:evtdisplay}
\end{figure}

\begin{figure}[!h]
\centerline{\includegraphics[height=7cm]{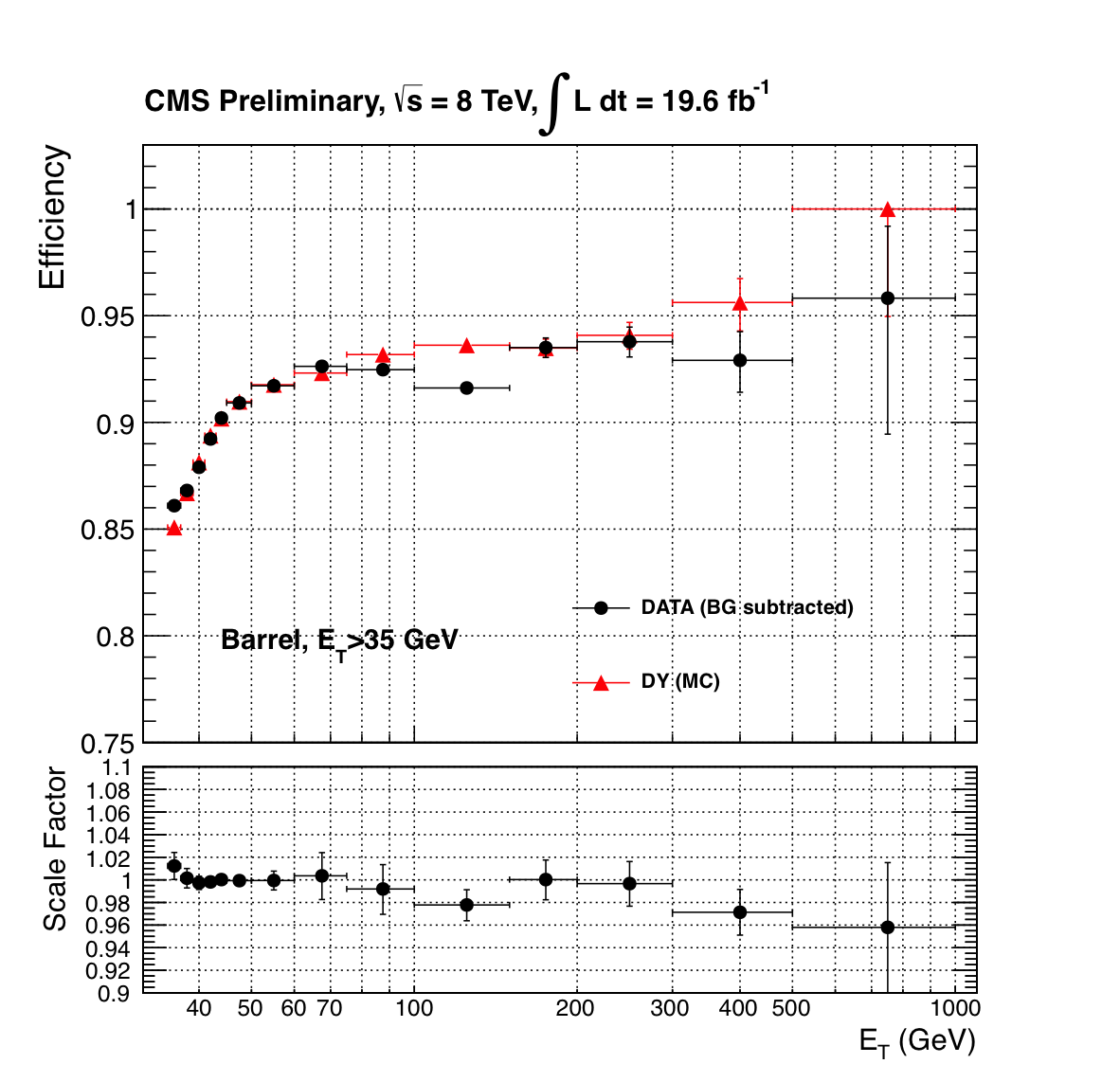}\includegraphics[height=7cm]{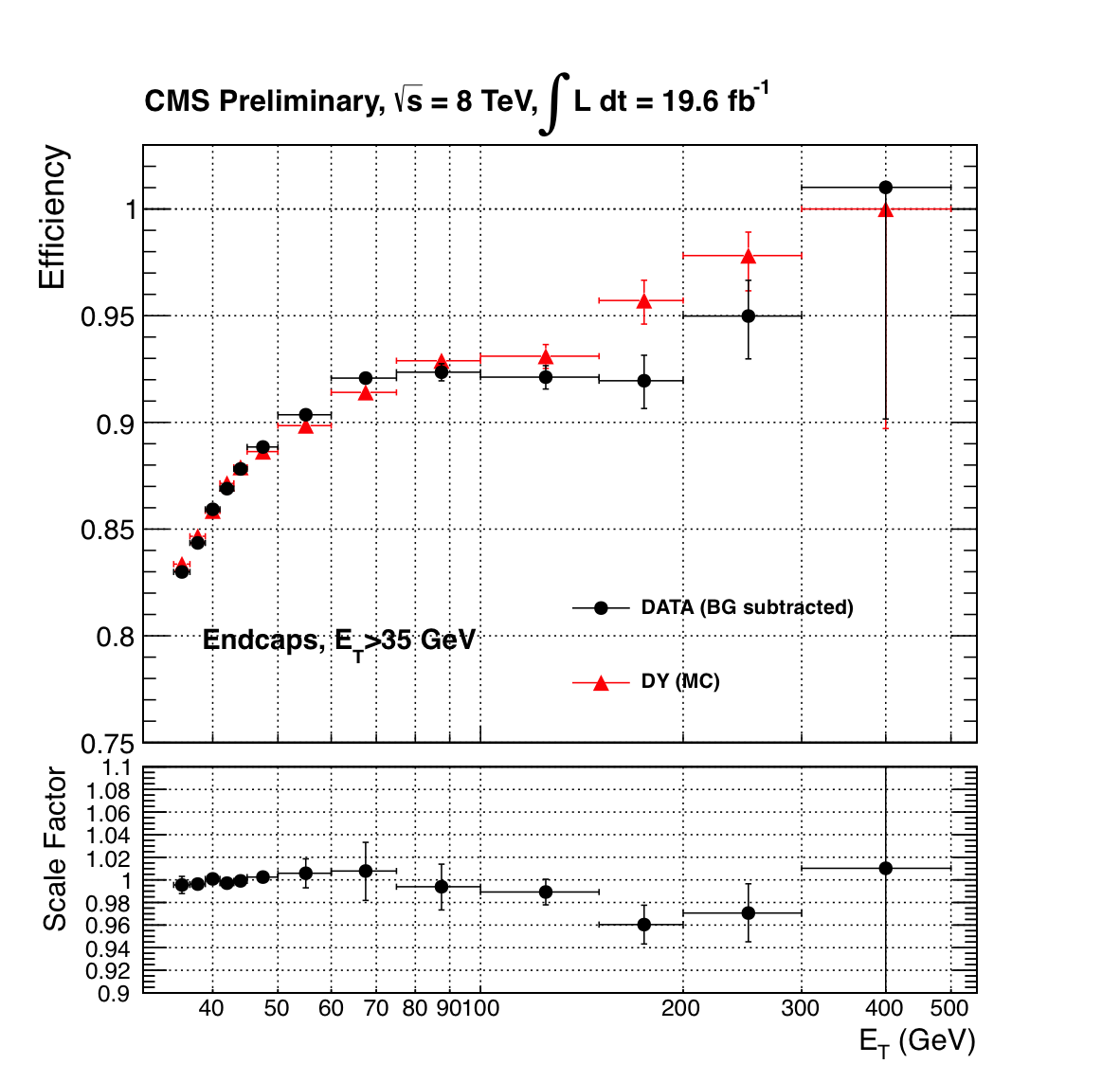}}
\caption{Efficiency of the electron selection used in this analysis as a function of $E_T$ for a simulated sample of Drell-Yan events (red triangles) and for the data after background subtraction (black dots).
The bottom part of the plot shows the scale factor defined as the ratio of data over simulation measurements. 
The results are shown separately for electrons emitted in the central (left) and forward (right) regions of the detector.
Only statistical errors are drawn for the efficiencies, while statistical error and systematical error on the background contamination are combined for the scale factor. 
}
\label{fig:effvspt}
\end{figure}

Only events with two high energy electrons candidates~\cite{ElectronRECO} are selected. 
Identification (e.g. shower shape, ratio of energy recorded in the HCAL and ECAL) and isolation (e.g. energy deposit in the calorimeters around the electron candidate) criteria are then applied.  
The set of requirements for each electron constitutes the High Energy Electron Pairs selection and is specifically optimized to be very efficient for a large range of the electron transverse energy $E_T$ (from $E_T$= 35 GeV to above 1 TeV).
A view of the highest mass event recorded by CMS in 2012 is shown in figure~\ref{fig:evtdisplay}.
As only very few events are expected in the high dielectron mass region, the selection efficiency is taken from simulation. 
It is therefore crucial to check that the simulation describes well the data.
This is done using the \textit{Tag and Probe} method. 
The principle is to select events with two real electrons from the Drell-Yan process and, in order to reduce the background, strong criteria are applied on one of the electron (the Tag). 
The efficiency of the selection under study is then measured on the second electron (the Probe). 
The resulting efficiency as a function of the electron transverse energy is shown in figure~\ref{fig:effvspt} for events at the Z pole where large statistics is available with low background. 
Within the uncertainties, a good agreement is obtained up to $E_T$ of several hundreds of GeV.

\section{Background}
\begin{figure}[h]
\centerline{\includegraphics[height=5cm]{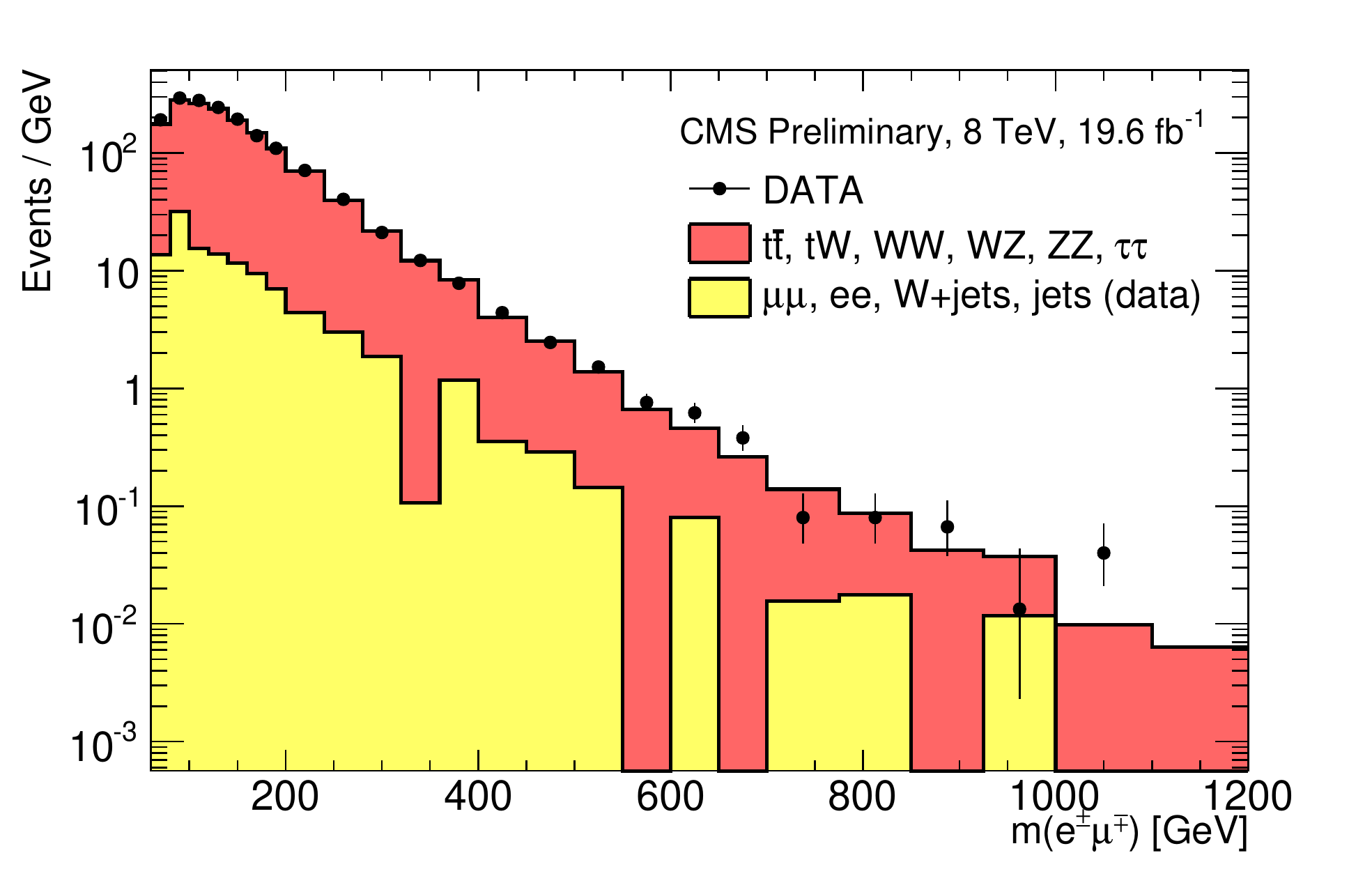}}
\caption{Invariant mass spectrum for $e\mu$ pairs. 
This distribution is used to validate the contributions of the $t\bar{t}$ and diboson processes in the dielectron or dimuon mass spectrum that are estimated from simulations.}
\label{fig:emumass}
\end{figure}

The main background to the signal is the Drell-Yan process, it is irreducible and is therefore estimated from simulations.
The second background comes from the leptonic decay of $t\bar{t}$ or diboson pairs. 
Simulations are used to estimate its contribution and are validated by comparing in data and simulations the invariant mass spectrum of $e\mu$ pairs (figure~\ref{fig:emumass}) for which those processes are dominant. 
After correcting for the difference in selection efficiency and acceptance between electron and muon, the number of $ee$ or $\mu\mu$ events from these processes is expected to be half the number of events in the $e\mu$ spectrum.  
Finally, there is also a small contribution from processes where at least one jet fakes an electron in the detector (multijets, W+jets and photon+jets).
This faking probability is measured directly from the data and allows to estimate the contribution to the final mass spectrum without using simulation. \par
The overall background estimation is checked with the data in the dielectron mass tail between 120 and 200 GeV/c$^2$, where no new physics is expected from previous experiments.

\section{Results}
\begin{figure}[h]
\centerline{\includegraphics[height=5cm]{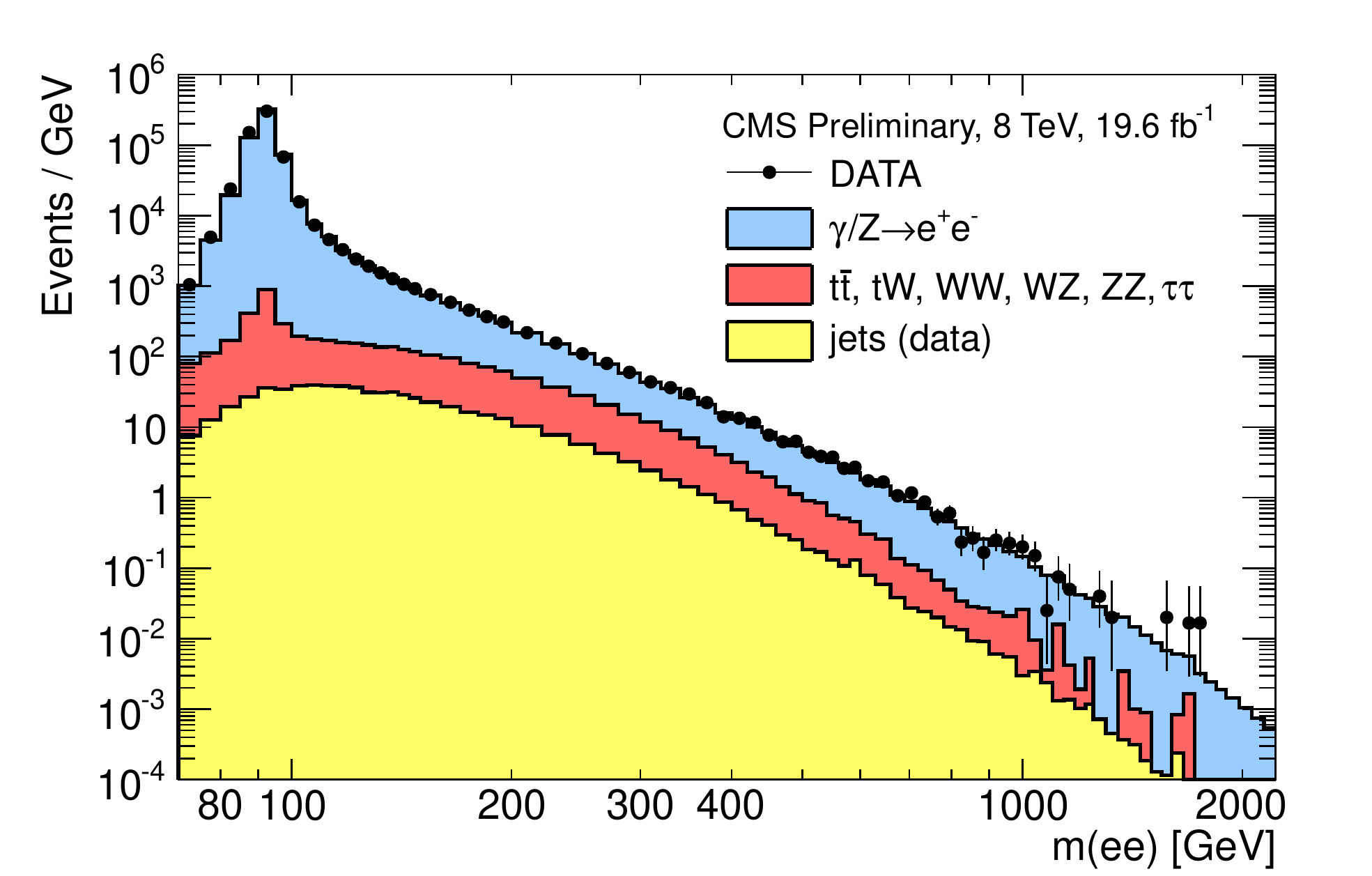}\includegraphics[height=5cm]{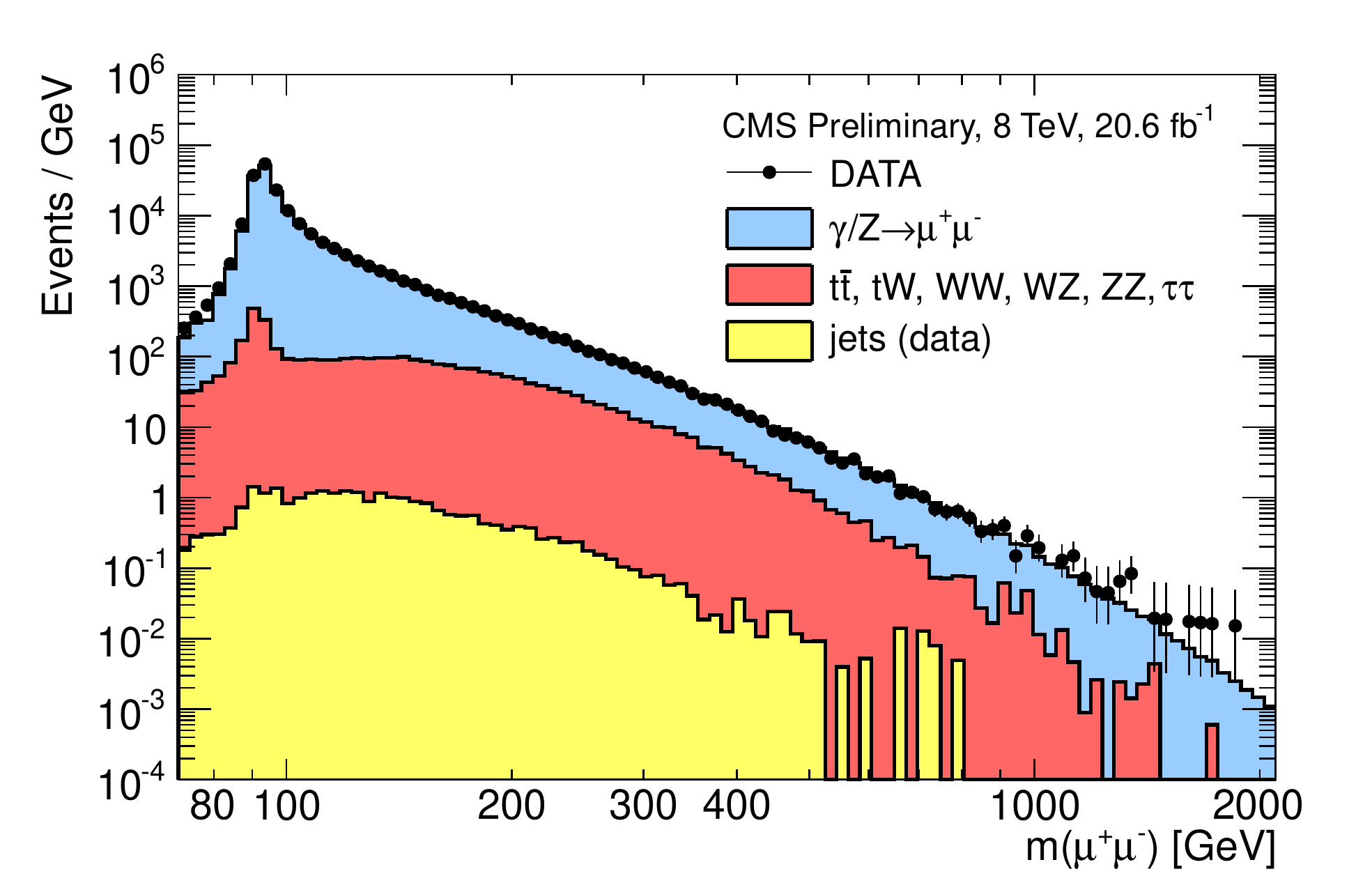}}
\caption{Final dielectron (left) and dimuon (right) mass spectrum obtained from the analysis of the complete 2012 dataset recorded by CMS.}
\label{fig:heepheepmass}
\end{figure}

\begin{figure}[h]
\centerline{\includegraphics[height=5.5cm]{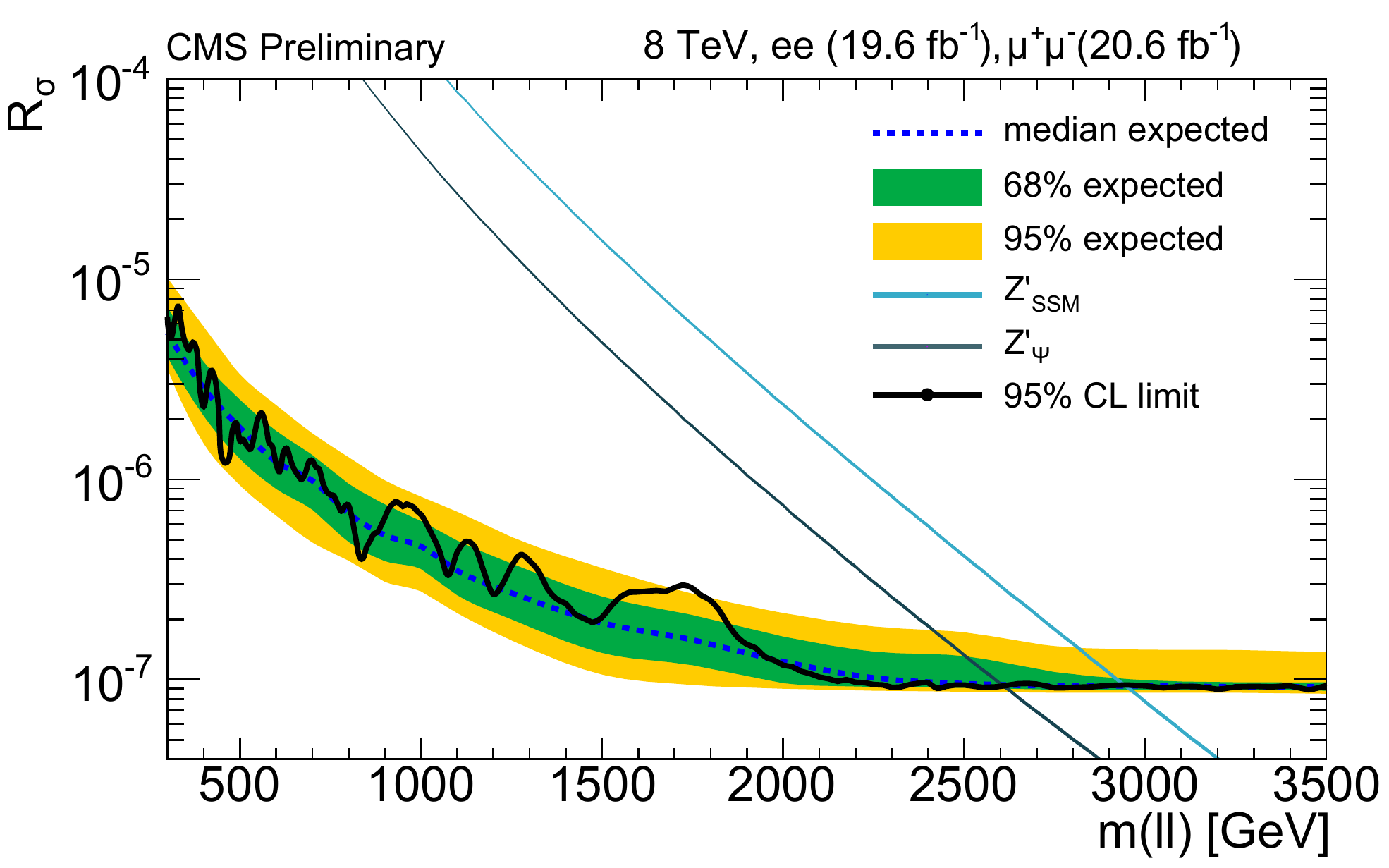}}
\caption{Observed (black line) and expected (green and yellow bands) 95\% C.L. upper limit on $R_\sigma$ as a function of the resonance mass after combination of the dielectron and dimuon channels. 
Also shown are the expected cross section curves for several benchmark scenarios.}
\label{fig:limits}
\end{figure}

The final mass spectra obtained with the complete 2012 dataset are shown in figure~\ref{fig:heepheepmass} for the dielectron (left) and dimuon (right) channels.
A good agreement between data and background is observed in the whole dilepton mass range. 
As no evidence of new physics is seen, generic upper limits on the cross section production of new spin 1 resonances are set. 
The variable under study is the ratio ($R_\sigma$) of the production cross section times branching ratio to two leptons for a new heavy boson and the Z boson. 
Results are shown in figure~\ref{fig:limits}; the limits are computed using an extended unbinned likelihood function. 
The width of the peak is assumed to be dominated by the detector resolution and the signal pdf is taken to be a gaussian. 
The background shape is fitted to the background estimate in the range $200<M_{l\bar{l}}<3500$ GeV/c$^2$. 
The upper limits on the production cross section are converted into lower limits on the mass of new resonances for specific models using their predicted cross section for events in a window of $\pm$ 40\% of the resonance mass following a theoretical prescription~\cite{LimitsCut}. 
The $Z'_{SSM}$ which has the same coupling than the Z boson and the $Z'_{\psi}$ predicted by E6 GUT model are excluded below 2.96 and 2.60 TeV/c$^2$ respectively. 
\section*{Acknowledgments}
The author is supported by the FNRS (Belgium) and by the Belgian Science Policy office (IAP VII/37).

\end{document}